\documentclass{article}

\usepackage{microtype}
\usepackage{graphicx}
\usepackage{subcaption}
\usepackage{booktabs} 
\usepackage{bm}
\usepackage{amsmath}
\usepackage{amssymb}
\usepackage{xcolor}

\usepackage{hyperref}

\newcommand{\mb}[1]{\mathbf{#1}}
\newcommand{\bs}[1]{\bm{#1}} 
\usepackage[accepted]{icml2019}

\icmltitlerunning{Bayesian Nonparametric Adaptive Spectral Density Estimation for Financial Time Series}

\begin{document}

\twocolumn[
\icmltitle{Bayesian Nonparametric Adaptive Spectral Density Estimation \\
           for Financial Time Series }




\begin{icmlauthorlist}
\icmlauthor{Nick James}{1, 3}
\icmlauthor{Roman Marchant}{1}
\icmlauthor{Richard Gerlach}{2}
\icmlauthor{Sally Cripps}{1,3}
\end{icmlauthorlist}

\icmlaffiliation{1}{Centre for Translational Data Science, Sydney, Australia}
\icmlaffiliation{2}{University of Sydney Business School, Sydney, Australia}
\icmlaffiliation{3}{School of Mathematics and Statistics, University of Sydney, Sydney, Australia}

\icmlcorrespondingauthor{Nick James}{nicholas.james@sydney.edu.au}

\icmlkeywords{Bayesian Nonparametrics, Spectral Density Estimation, Reversible Jump Markov Chain Monte Carlo, Financial Time Series}

\vskip 0.3in
]




\begin{abstract}
 Discrimination between non-stationarity and long-range dependency is a difficult and long-standing issue in modelling financial time series. This paper uses an adaptive spectral technique which jointly models the non-stationarity and dependency of financial time series in a non-parametric fashion assuming that the time series consists of a finite, but unknown number, of locally stationary processes, the locations of which are also unknown. The model allows a non-parametric estimate of the dependency structure by modelling the auto-covariance function in the spectral domain.  All our estimates are made within a Bayesian framework where we use a \emph{Reversible Jump Markov Chain Monte Carlo} (RJMCMC) algorithm for inference. We study the frequentist properties of our estimates via a simulation study, and present a novel way of generating time series data from a nonparametric spectrum.  Results indicate that our techniques perform well across a range of data generating processes. We apply our method to a number of real examples and our results indicate that several financial time series exhibit both long-range dependency and non-stationarity. 
\end{abstract}

\section{Introduction}
\label{submission}

Modelling the volatility of financial time series has been the subject of much interest since the deregulation of world financial markets, which began in the late 1970's. It is a difficult task. First financial time series are often non-stationary, by which we mean that the statistical properties change over time, making the development of statistical models problematic. Second, even stationary financial time series exhibit non-standard features such as  volatility clustering  \citep{Mandelbrot1963} and related kurtosis. Third, the signal to noise ratio is high making it difficult to detect any underlying trends. The primary contribution of this paper is to work in the spectral domain to capture and distinguish between features of time series data, such as non-stationarity and long-range dependency and compare estimates of these features with estimates obtained using parametric time domain models.

Models for the time-varying nature of volatility in financial markets began with \citet{Engle1982}, who introduced the \emph{AutoRegressive Conditionally Heteroskedastic} (ARCH) model.  \citet{Bollerslev1986} extended this model to the more parsimonious \emph{Generalized ARCH} (GARCH), while \citet{Taylor1982, Taylor1986, Taylor1986_2} developed the \emph{Stochastic Volatility} (SV) framework over the same period. These first generation volatility models are conditionally Gaussian, with the dynamic volatility component meant to account for the leptokurtosis present in most financial return series.  \citet{Bollerslev1987} allowed for conditionally Student-t returns in a GARCH process, specifically increasing the level of this aspect able to be captured. Various salient features of observed financial returns, e.g. the leverage effect, whereby volatility is higher in falling, compared to rising, markets, and the non-stationary aspects, including the time-varying nature of the conditional distribution of returns and possible structural break components, are allowed for in subsequent extended GARCH models in the literature. These include the EGARCH \citep{Nelson1991}, GJR-GARCH \citep{Glosten1993}, T-GARCH \citep{Zakoian1994} and T-SV \citep{So2002} models, all attempting to capture the leverage effect; and via \emph{Markov Switching GARCH} (MS-GARCH) \citep{Cai1994, Hamilton1994, Gray1996, Haas2004} and MS-SV models \citep{Lam1998}.

Bayesian estimation has been prominent in this area, especially for SV models where the likelihood, without conditioning on the latent stochastic process, does not exist in closed form and simulation-based and/or data augmentation methods, including \emph{Markov Chain Monte Carlo} (MCMC), are standard. As high frequency data became more and more available, volatility modelling moved first to directly model realized measures, such as realized variance \citep{Andersen2003}, and then to extensions of GARCH and SV processes, allowing realized measures as inputs that drive volatility changes in the model, e.g. the GARCH-X model of \citet{Hwang2007}. More recently, \citet{Hansen2011} developed the realized GARCH framework, allowing an extra measurement equation capturing the contemporaneous relationship between the latent volatility and the realized measure.  All these models make assumptions about distribution of the noise and have parametric representations of the evolution of the volatility and although some methods may explicitly model regime shifts and stochastic behaviour, if the parametric form of the model does not  resemble  the  underlying  phenomenology  of  the data generation process it will perform poorly.  

Flexibly estimating the time-dependency of a phenomenon via the spectral density goes back to the 1950's \citep{Whittle1957}. However, it is not often applied to financial time series, despite several appealing reasons for doing so as pointed out by \citet{Chaudhuri2015}.  First, studying frequency components of security's return processes can provide insight into previously unseen economic structure driving price movements. Secondly, as investment time horizons can range from microseconds to many years, time-specific risks can be accounted for in portfolio construction decisions. Thirdly, frequency domain analysis can also compare strategies that operate on different timescales, and may provide diversification across investment strategies operating on varying timescales. Finally, frequency-domain measurements offer more understandable representations of the complex periodic dynamics financial markets may exhibit.

In addition to these advantages, latest developments in spectral analysis, in particular the concept of local stationarity  developed by \citet{Dahlhaus1997} and built on by \citet{Rosen2009,Rosen2012}, have led to the development of flexible nonparametric methods for estimating time varying spectra.

We use the technique of \citet{Rosen2012} to jointly estimate a time-varying non-parametric spectrum for financial time series data, and to distinguish between non-stationarity and long-range dependency, as evidenced by volatility clustering. We use this flexible time-varying spectrum to simulate "ground truth" data in the spectral domain and convert back into time domain. This enables us  to compare time domain models of volatility with each other and with spectral techniques. All our parameter estimates are made within a Bayesian framework where we use a MCMC algorithm for inference. Rather than modelling volatility with a conditionally stationary process as proposed by ARCH/GARCH/SV-style models, we assume that the data generating process is non-stationary, and consists of an unknown number  and location of locally stationary processes. 

The remainder of this paper is organised as follows. Section \ref{sec:model_prior} describes the model, priors and estimation procedure. Section \ref{sec:experiments} shows validation over simulated and real-world data. Finally, Section \ref{sec:conclusion} draws conclusions from our experiments.

\section{Model, Priors and Estimation}
\label{sec:model_prior}
\subsection{Model for Non-stationary processes}
Suppose $\{y_t\}_{t=1,\ldots,T}$ is a time series with observations from a Dahlhaus locally stationary process with evolutionary spectrum $f(\nu,t)$, which we wish to estimate. To do this, we assume that the time series  consists of $K$ piecewise stationary processes, each of length $n_s$ for $s=1,\ldots,K$. Given a partition of $K$ segments, we define the partition points to be  $\xi_K=(\xi_{0,K},\xi_{1,K}\ldots \xi_{K,K})$, with $\xi_{0,K}=0$ and $\xi_{K,K}=T$ so that the set $A_s$ is given by $A_s=\{t;\xi_{s-1}+1 <t<\xi_s\}$ as in   \cite{Rosen2012}.  Therefore, we can rewrite
\begin{equation}
y_{t}=\sum_{s=1}^K y^{(s)}_t\delta(t,A_{s,K})
\end{equation}
where, $\delta(t,A_{s,K})=1$, if $t\in A_s$ and $\delta(t,A_{s,K})=0$ otherwise, and where the $y^{(s)}_{t}$'s are independent stationary processes, for $s=1,\ldots,K$, each with spectral density $f_{s,K}(\nu)$.

The joint probability density function of a realization $\mb y=(y_1,\ldots,y_T)$ given  the individual spectra $\mb F_K=(\mb f_{1,K}(\nu),\ldots,\mb f_{K,K}(\nu))$, the number of segments $K$, and the partition points $\bs \xi_K$ is
\begin{align}
\label{eqn_1}
p(\mb y|\mb F_K, K,\bs\xi_K)= &\nonumber\\
\prod_{s=1}^{K}&p\left(y_{\xi_{(s-1,K)}+1},\ldots,y_{\xi_{(s,K)}}|\mb f_{s,K}(\nu)\right)
\end{align}

\subsection{Priors}
\subsubsection{ Prior for Spectra}
Given a partition defined by $K$ segments and their respective parition points $\bs\xi_K$, and a realization $\mb y^{s)}$, our goal is to estimate the unknown spectra $f_{s,K}(\nu)$, for $\nu\in(0.5,0.5)$. To motivate a prior for $f_{s,K}(\nu)$, we frame the problem of estimating the autocovariance structure of a time series, given by the spectrum, as a nonparametric regression estimation problem. In effect turning a covariance estimation problem into a  mean estimation problem, which is more parsimonious and tractable.

To elaborate, define the \emph{Discrete Fourier Transform} (DFT) for segment $s$ of length $n_s$, at frequency $\nu_k$ to be 
\begin{align}
x_s(\nu_k) = \frac{1}{\sqrt{n_s}}\sum_{t=1}^{n_s}&\,\,y_{t+\xi_{s-1}+1} \times\\
&\left(\cos(2\pi\nu_kt)-i\sin(2\pi \nu_kt)\right)\nonumber\,\,,
\end{align}
where $\nu_k=k/n_s \,\, \forall k \in \{0,1,\ldots,(n_s-1)\}$. Let the periodogram at frequency $\nu_k$, $I(\nu_k)$, be the squared modulus of the DFT
\begin{equation}
I_s(\nu_k)=\big|x_s(\nu_k)\bar{x}_s(\nu_k)\big| \,\,.
\end{equation}
Then Whittle \cite{Whittle1957} showed that the distribution of $\mb x_s=\left(x_s(\nu_1)\ldots,x_s(\nu_{n_s})\right)$, under certain regularity conditions, is complex normal so that
\begin{equation}
\label{eq:whittle}
\mb x_s \sim \prod_{k=1}^{n_s}\frac{1}{\pi f_s(\nu_k)}\exp\left(-\frac{I_s(\nu_k)}{f_s(\nu_k)}\right).
\end{equation}
This representation suggests that the $I_s(\nu_k)$ are i.i.d. with $I_s(\nu_k)\sim \exp(f_s(\nu_k))$ and therefore
\begin{equation}
\log(I_s(\nu_k))=\log(f_s(\nu_k))+\epsilon_k;\;\epsilon_k\sim \log(\exp(1))
\end{equation}
Letting $w_s(\nu_{k}) = \log \left(I_s(\nu_{k})\right)$ and $g_s(\nu_{k}) = \log\left(f_s(\nu_{k})\right)$ we have
\begin{equation}
w_{s}(\nu_{k}) = g_s(\nu_{k}) + \epsilon_{k},
\end{equation}
To place a prior on the unknown function $g_s(\nu_{k})$ we decompose it into its linear and non-linear components so that $g_s(\nu_{k})= \alpha_{s0}  + h_s(\nu_{k})$  and place a Gaussian Process prior  over the unknown function $h_s(\nu_{k})$, see for example \cite{Wahba1990}. Specifically we assume
\begin{equation}
h_s(\nu_k) = \tau_s W(\nu_k)
\end{equation}
or equivalently,
\begin{equation}
\mb h_s  = \left(h_s(\nu_1),\ldots,h_s(\nu_{n_s})\right)
          \sim \mathcal{N}\left(0,\tau_s^2\Omega\right)
\end{equation}
where $W(.)$ is a Wiener process, $\tau_s^{2}$ is a smoothing parameter and the $i^{th}$, $j^{th}$ element of $\Omega$, $\omega_{ij} = \mbox{cov}(h_s(\nu_i),h_s(\nu_j))=\min(\nu_{i},\nu_{j})$. 

For computational convenience we write $\mb h_s$ as a linear combination of basis functions by performing an eigenvalue decomposition on $\Omega=QDQ'$. Specifically we let $X=QD^{1/2}$ be the design matrix and  $\bs\beta_s\sim(0,\tau^2_sI_{n_s})$ be the vector of regression coefficients, so that $\mb h_s=X\beta_s$ has the required distribution.
We follow Wood et al. \cite{Wood2002} and Rosen et al. \cite{Rosen2009} and keep only those basis functions corresponding to the 30 largest eigenvalues, for computational speed.
\subsubsection{Prior for Partition}
The partition is defined by the number of of locally stationary segments $K$ and the partition points, $\bs\xi_K$, given $K$.
The prior on the partition $\Pr(K,\bs\xi_S)=\Pr(\bs\xi_s|K)\Pr(K)$ $\bs\xi_{s,K}$ is as follows;
\begin{equation}
\Pr(K)=\frac{1}{S}
\end{equation}
where $S$ is the the upper limit for the number of segments, in the experiments which follow this is typically set to be 30. Given $K$ we decompose the prior on $\bs\xi_K$ into a sequence of discrete uniform priors,so that 
\begin{equation}
\Pr(\bs{\xi_{K}} \mid K) = \prod^{K-1}_{s=1} \Pr(\xi_{s,K}|\xi_{s-1}, K)\,\,,
\label{part_prior}
\end{equation}
where $\Pr(\xi_{j,m}=t \mid m)$ = $1/p_{s,K},$ for $s = 1,\ldots,K - 1,$  $p_{s,K}$ is the number of available locations for partition point $\xi_{s,K}$ and is equal to
$T - \xi_{s-1,K} - (K-s + 1)t_{\min} + 1$. The quantity $t_{\min}$ is a user chosen number.  It represents the minimum number of observations that are deemed sufficient for the Whittle likelihood approximation to hold.  In this paper we set this to be 50, however we note that this is arbitrary, and indeed there is a substantial literature which discusses  the quality of the Whittle approximation.

The prior in Equation~\ref{part_prior} states that the first partition point is equally likely to occur at any
point in the time series subject to the constraint that there
are at least $t_{\min}$ observations in each of the $K$ segments. The
prior on subsequent partition points is similar and states
that, conditional on the previous partition point, the next
partition point is equally likely to occur in any available
location, again subject to the same constraint see \cite{Rosen2012} for details.

\subsection{Generation of Temporal Data}
\label{sec:generate_time}
A contribution of this paper is to use the time-varying spectra estimated as in \cite{Rosen2012} to generate a time series, without assuming the time domain data generating process.  This is achieved using the result that the DFT's of the realization of a process, are approximately normally distributed if the joint cumulants of that process, of orders greater than 2, are absolutely summable \cite{Brillinger1975}.

Let
\begin{eqnarray*}
\mathbf{x}_{r} &= &(x_{(0,r)},...,x_{(n_s-1,r)})\,\,, \\
\mathbf{x}_{i} &= &(x_{(0,i)},...,x_{(n_s-1,i)})\,\,,
\end{eqnarray*}
be the real and imaginary components of the DFT for a set of realizations from a locally stationary process $s$ of length $n_s$. The distribution of these quantities for a zero-mean process are;
\begin{eqnarray*}
x_{\left(0,r\right)} &\sim &\mathcal{N}(0, f_s(\nu_0))\\
x_{\left(0,i\right)} &\sim& \delta(0) \,\\
x_{\left(1:\frac{n_s}{2}-1,r\right)} &\sim& \mathcal{N}\left(0,f_s(\nu_{1:\frac{n_s}{2}-1)}/2\right)\,\\
x_{\left(1:\frac{n_s}{2}-1,i\right)} &\sim& \mathcal{N}\left(0,f_s(\nu_{1:\frac{n_s}{2}-1)}/2\right)\,\\
\end{eqnarray*}
where $\delta(.)$ is the Dirac delta function. If $n$ is even then
\begin{eqnarray*}
x_{\left(\frac{n_s}{2},r\right)} &\sim &\mathcal{N}\left(0, f(\nu_{n_s/2})\right)\\
x_{\left(\frac{n_s}{2},i\right)} &\sim& \delta(0) \,\,.
\end{eqnarray*}
To ensure symmetry we set
\begin{eqnarray*}
x_{\left(\frac{n}{2}+1:n-1,r\right)} &= x_{\left(\frac{n}{2}-1:1,r\right)}\,\,\\
x_{\left(\frac{n}{2}+1:n-1,i\right)} &= - x_{\left(\frac{n}{2}-1:1,i\right)}\,\,,
\end{eqnarray*}
So that given a time-varying spectrum $f(\nu,t)$, for $\xi_{s-1,S}<t\le\xi_{s,S}$ we generate $\mathbf{x}_{r}$ and $\mathbf{x}_{i}$ and form $\mb x=\mb{x}_{r}+i \mb{x}_{i}$ and apply the Inverse-DFT to generate the time series corresponding to each locally stationary process and so obtain a time domain realization from a non-stationary process.

\subsection{Estimation}
In this paper we take a Bayesian approach and estimate the unknown time-varying spectrum by its posterior mean 
\begin{eqnarray*}
\mathbb{E}[f(\nu,t)|\mb y] = \sum_{K=1}^S\sum_{j=1}^{p^{(K,T)}}\left\{f(\nu,t)|\mb y,K,\bs\xi_K)\right\}\\
\times\Pr(\mb\xi_S|K,\mb y)\Pr(K|\mb y)
\end{eqnarray*}
where the sum is over all possible partitions and
\begin{align}
\mathbb{E}\left[f(\nu,t)|\mb y,K,\bs\xi_K\right] = &\\ \int \mathbb{E}[f(\nu,t)&|\mb y, K,\bs\xi_S, \mb F_K]p(\mb F_K|\mb y,K,\bs\xi_K)d\mb F_K \,\,. \nonumber
\end{align}
We use \emph{Reversible Jump MCMC} (RJMCMC) to perform the required transdimensional integration, see \cite{Rosen2012} for details.

\section{Experiments}
\label{sec:experiments}
This section validates the use of more flexible, adaptive non-parametric models for estimating spectrum of financial time series and its volatility. The experiment setup is as follows, we evaluate the goodness of fit for different techniques over data with a known generative process and over real-world data from the daily returns and squared returns of the NASDAQ Index from 2002-2018 and the GBP:USD from 2010-2018. Section \ref{sec:exp_simulated_data} presents details on the data generation processes and evaluation of results for a known  time-varying spectral density. Section \ref{sec:experiment_real_data} shows the results of fitting different models over the returns and squared returns of the NASDAQ and GBP:USD. 

\subsection{Simulated Data}
\label{sec:exp_simulated_data}

\begin{figure*}[ht]
\centering
\begin{subfigure}{0.32\textwidth}
\includegraphics[width=\columnwidth]{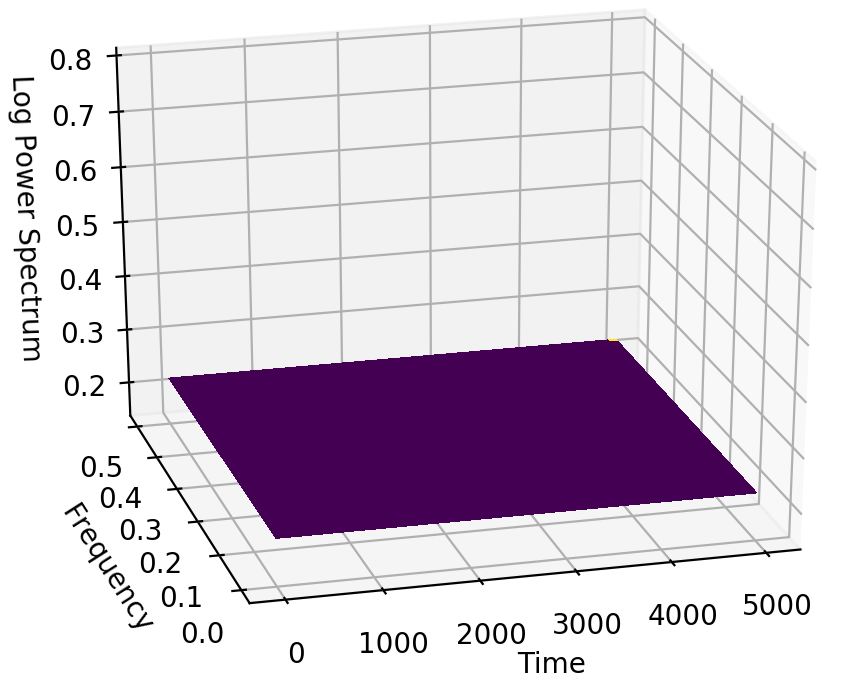}
\caption{GARCH Log Spectrum}
\label{fig:garch_spectrum}
\end{subfigure}
\begin{subfigure}{0.32\textwidth}
\includegraphics[width=\columnwidth]{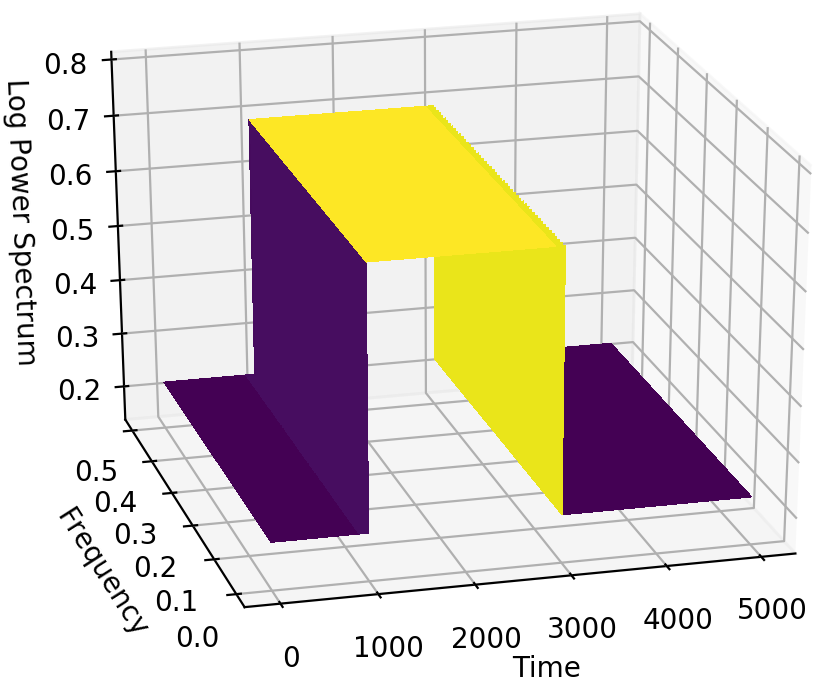}
\caption{Regime Log Spectrum}
\label{fig:regime_spectrum}
\end{subfigure}
\begin{subfigure}{0.32\textwidth}
\includegraphics[width=\columnwidth]{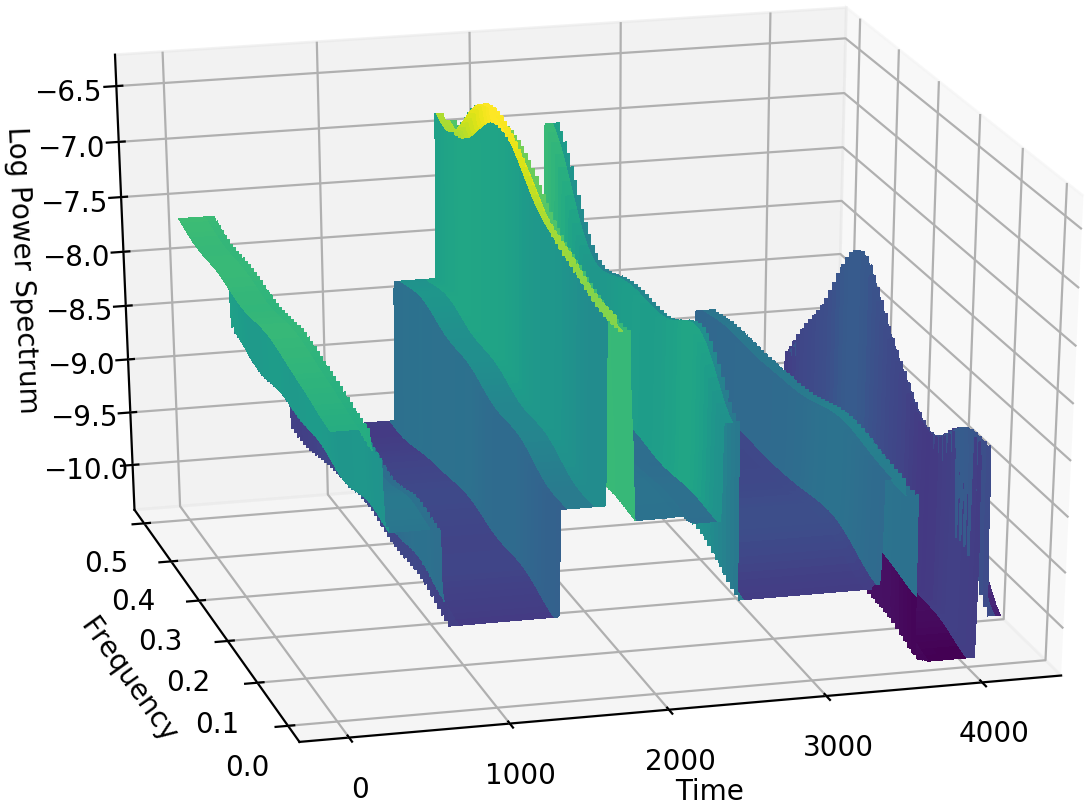}
\caption{AdaptSpec Log Spectrum}
\label{fig:adaptspec_spectrum}
\end{subfigure}
\begin{subfigure}{0.32\textwidth}
\includegraphics[width=\columnwidth]{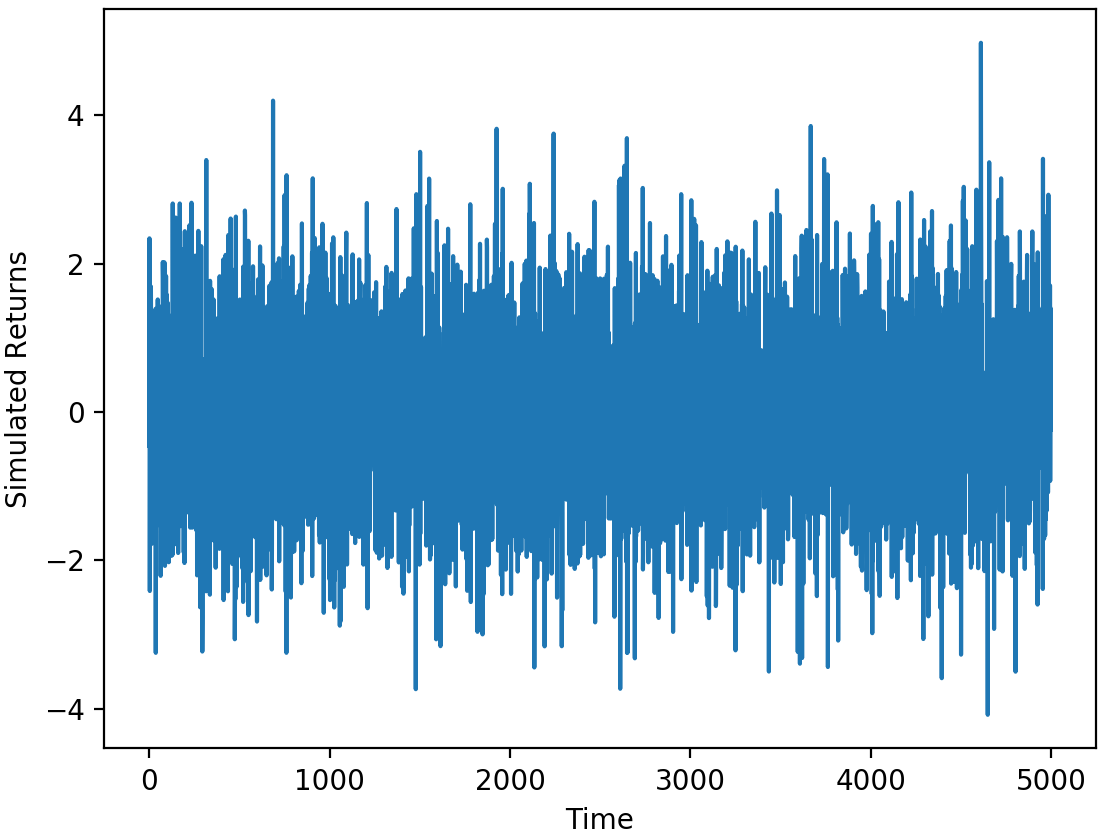}
\caption{GARCH Time Series}
\label{fig:garch_time}
\end{subfigure}
\begin{subfigure}{0.32\textwidth}
\includegraphics[width=\columnwidth]{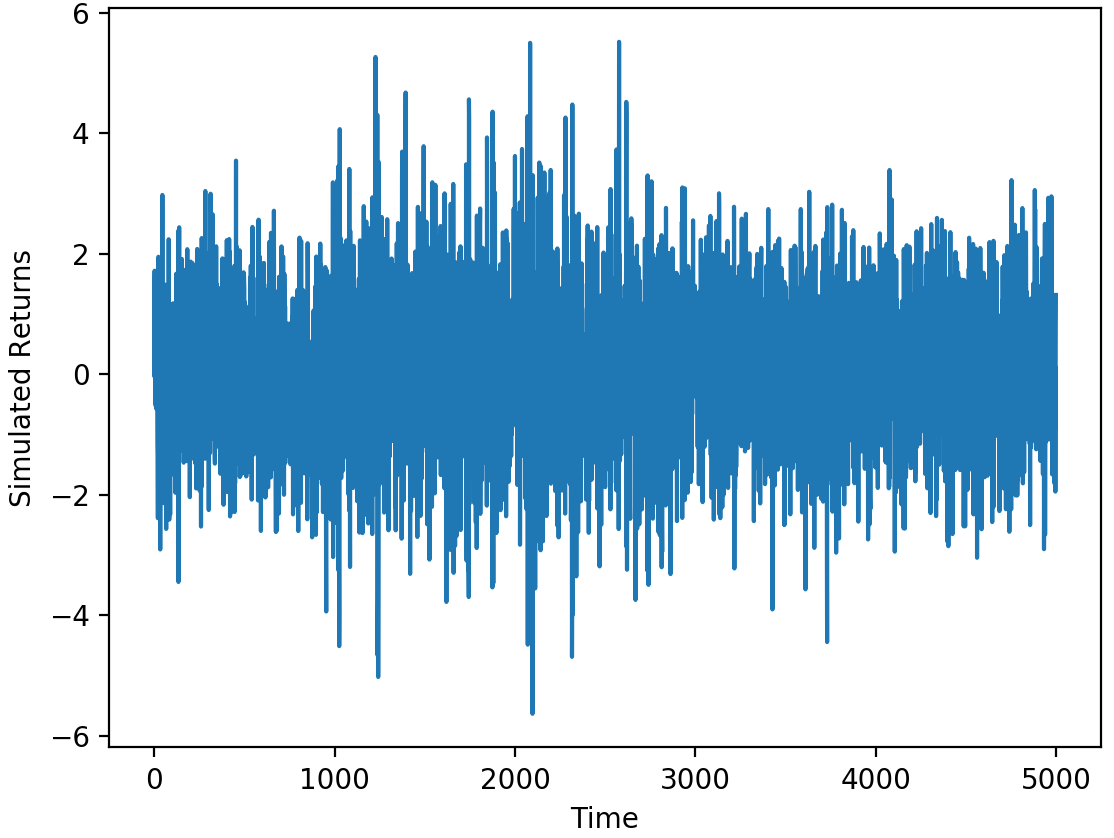}
\caption{Regime Time Series}
\label{fig:regime_time}
\end{subfigure}
\begin{subfigure}{0.32\textwidth}
\includegraphics[width=\columnwidth]{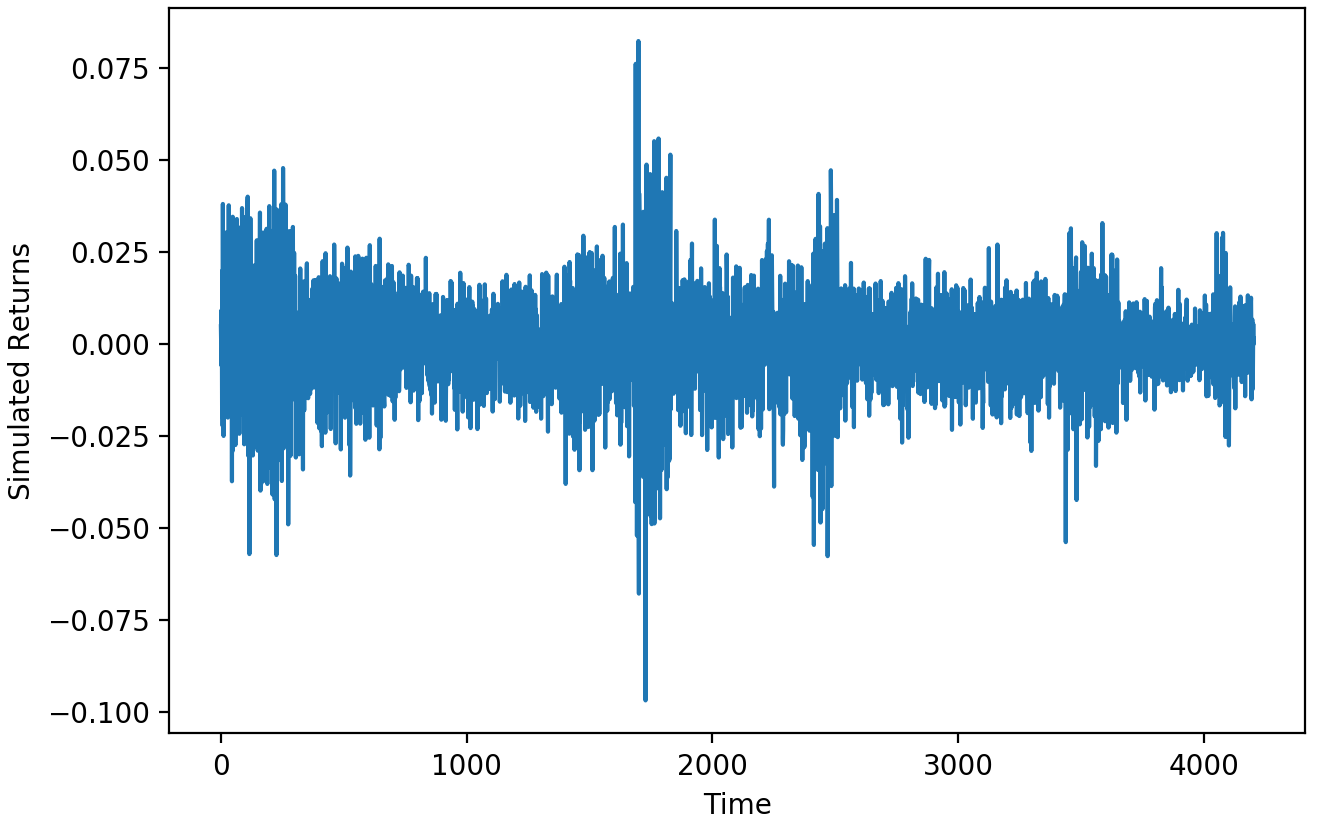}
\caption{AdaptSpec Time Series}
\label{fig:adaptspec_time}
\end{subfigure}
\caption{Spectra and example realisations.}
\label{fig:simulation_data}
\end{figure*}

To compare the performance of various models for financial returns and volatility, in terms of the ability of the model to recover the true data generating process, we simulate data using three models for time series. The first model is a stationary process, while the second and third models are non-stationary processes. 
The first model is a GARCH (1,1) process. The second model is a regime-switching GARCH (1,1) process \cite{Haas2004, Ardia2016} and the third is the AdaptSpec model of \citet{Rosen2012}. The data generating process of a  GARCH(1,1) model \ref{fig:garch_time} is given by
\begin{align}
y_t&\sim \mathcal{N}(\mu,\sigma^2_t)\\
    \sigma^2_t & = \alpha_{0} + \alpha_{1} \eta_{t-i}^{2} + \beta_{1}\sigma^2_{t-1},
\end{align}
where $\eta_{t-1}=y_{t-1}-\mu$. We set $\mu=0, \alpha_0=1, \alpha_1=0.1, \beta_1=0.1.$, so that the process is stationary with an unconditional variance, $\sigma^2_{uc}=\frac{\alpha_0}{(1-\alpha_1-\beta_1)}$. Figure \ref{fig:garch_spectrum} shows a sample spectrum and Figure \ref{fig:garch_time} the associated realisation in time.

The second model we generate data from is a Regime-Switching GARCH model \ref{fig:regime_time} as in \cite{Ardia2016, Haas2004}.  Specifying a model which allows for regime-switching is one way of accounting for non-stationarity. For each point in time $t$, a latent state variable $s_{t}$ for $t \in \{1,2,..,T \}$, determines the regime from which the observation is generated. Let $\Pr(s_{t}=j \vert \mb y)$ be the probability that an observation at time $t$ was generated by regime $j$, for $j = 1,\ldots,N_{R}$, where $N_{R}$ is the number ofpossible regimes. Our Regime-Switching model is the following \cite{Bauwens2014, Haas2004} 
\begin{align}
y_t \vert s_{t}\!\!=\!\!j &\sim N\left(\mu_{j}, \sigma_{jt}^{2}\right) \\
\sigma_{jt}^{2} \vert s_{t}\!\!=\!\!j &= \alpha_{0,j} + \beta_{1,j} \sigma_{t-1}^{2} + \alpha_{1,j} \eta_{t-1}^{2}
\end{align}
For our simulation we set $N_{R} = 2$. Define $K_{R}$ to be the number of segments generated by the $N_{R}$
regimes, so that $K_{R} \ge N_{R}$. The location of the regime switches are defined by the cutpoints 
$\mb c=(c_1,\ldots,c_{K_R})$. Let $\mb r=(r_1,\ldots,r_{K_R})$
be an indicator vector denoting the regime which generates the data in segment $k$, so that  $r_k=j$, if segment $k$ was generated by regime $j$. 
For our simulation we set $K_{R} = 3$, $\mb c=(1000,3000,5000)$ and $\mb r=(1,2,1)$ . Our set of parameters in our regime switching model are, $\alpha_{0,1} = 1$,  $\alpha_{1,1} = 0.1$, $\beta_{1,1} = 0.1 $, $\alpha_{0,2} = 1 $, $\alpha_{1,2} = 0.3 $ and $\beta_{1,2} = 0.2 $.

The  third model for generating data is now described.  Using the model in \cite{Rosen2012} we obtained an estimate of the posterior mode of the number of locally stationary processes for the NASDAQ daily returns from 2002 to March 2018, denoted by $\hat{K}_{NAD}$ and an estimate of the posterior mean of the spectra\ref{fig:adaptspec_spectrum} corresponding to those locally stationary processes. We generated 50 realizations \ref{fig:adaptspec_time} of the real and imaginary components of the DFT's, $\mb x_{s,r}$ and $\mb x_{s,i}$ respectively each of length $n_{s,\hat{K}_{NAD}}$ , then the inverse-DFT was applied to obtain 50 time series, $\mb y_s$ each of length $n_{s,\hat{K}_{NAD}}$, for $s=1,\ldots,\hat{K}_{NAD}$ as described in Section \ref{sec:generate_time}.  These  $\hat{K}_{NAD}$ time series were concatenated, so that 50 realizations of a non-stationary process, of length $\sum_{s=1}n_{s\hat{K}_{NAD}}$, were obtained. 

In what follows we shall refer to these three data generating processes as GARCH, Regime and AdaptSpec.

\subsection {Metrics to measure performance}
 
To assess the relative performances of the GARCH, Regime, and the AdaptSpec models we use \emph{Mean Squared Error} (MSE) and \emph{Symmetric Kullback Liebler} (SKL) divergence. We define the quantities as follows
\begin{eqnarray*}
SKL &=& \sum_{t=1}^T \sum_{k=0}^{T-1} f(\nu_{k},t)\log \frac{f(\nu_{k},t)}{\hat{f}(\nu_{k},t)} \nonumber\\
&+& \hat{f}(\nu_{k},t)\log \frac{\hat{f}(\nu_{k},t)}{f(\nu_{k},t)} \\
MSE &=& \sum_{t=1}^T\sum_{k=1}^{n}(\hat{f}(\nu_k,t)-f(\nu_k,t))^2
\end{eqnarray*}
where $f(\nu,t)$ is the true time-varying spectrum and $\hat{f}(\nu,t)$ is an estimate of this true spectrum. In what follows we use  the subscripts $G$, $R$, and $AD$, to refer to the GARCH, Regime and AdaptSpec models respectively.
Plots of the true log spectra ${f}_G(\nu,t)$, ${f}_R(\nu,t)$ and ${f}_{AD}(\nu_t, t)$, used to generate the data along with an example of a realization appear in Figure \ref{fig:simulation_data}.

\begin{figure*}[ht]
\centering
\begin{subfigure}{0.32\textwidth}
\includegraphics[width=\columnwidth]{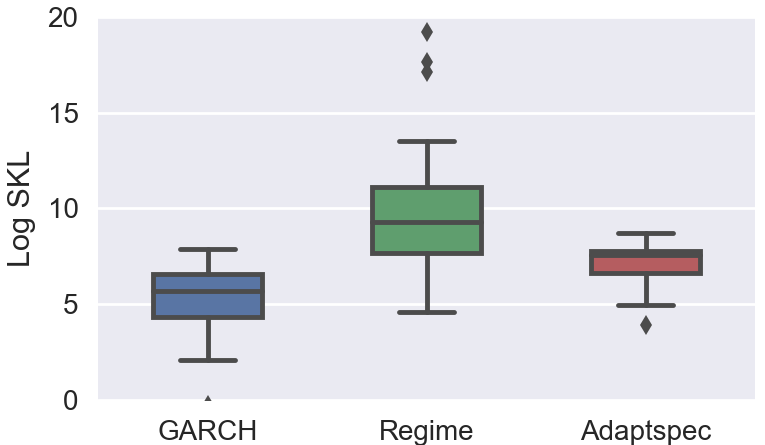}
\caption{GARCH Generated Data}
\label{fig:garch_skl}
\end{subfigure}
\begin{subfigure}{0.32\textwidth}
\includegraphics[width=\columnwidth]{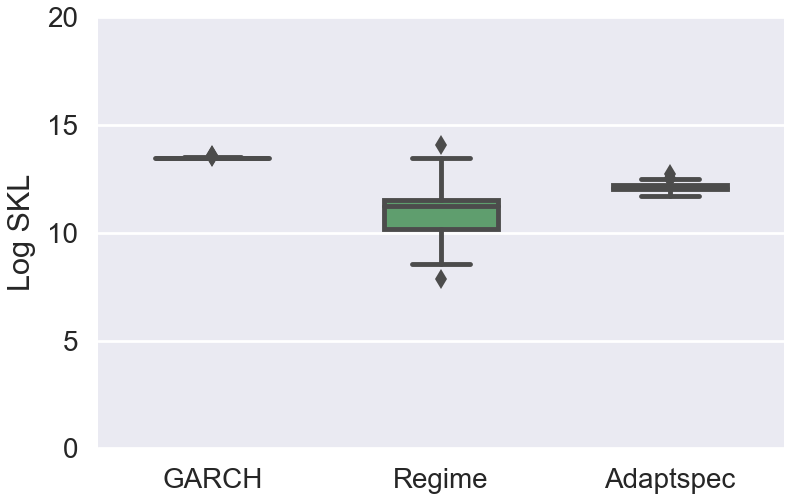}
\caption{Regime Generated Data}
\label{fig:regime_skl}
\end{subfigure}
\begin{subfigure}{0.32\textwidth}
\includegraphics[width=\columnwidth]{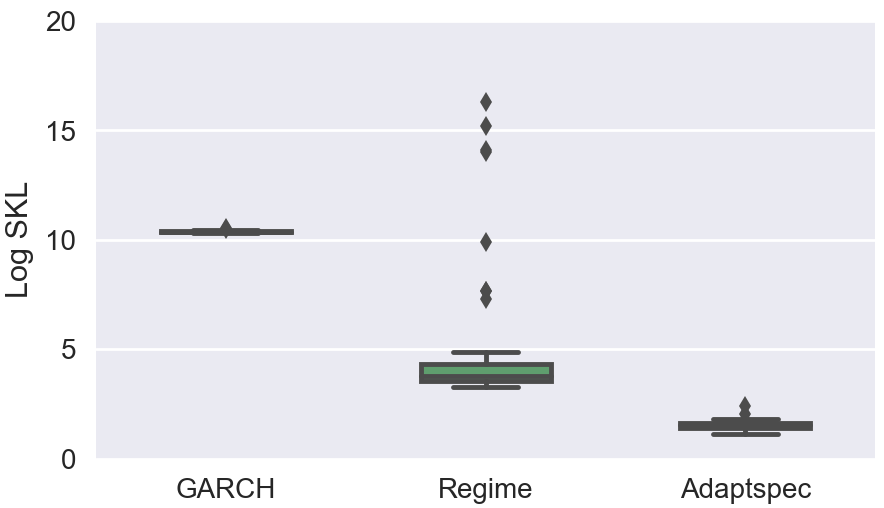}
\caption{AdaptSpec Generated Data}
\label{fig:adaptspec_skl}
\end{subfigure}
\caption{Boxplot of the $\log(SKL)$ divergence for three estimators from 50 realisations generated from each respective process.}
\label{fig:simulation_box_plots}
\end{figure*}

\begin{figure*}[ht]
\centering
\begin{subfigure}{0.32\textwidth}
\includegraphics[width=\columnwidth]{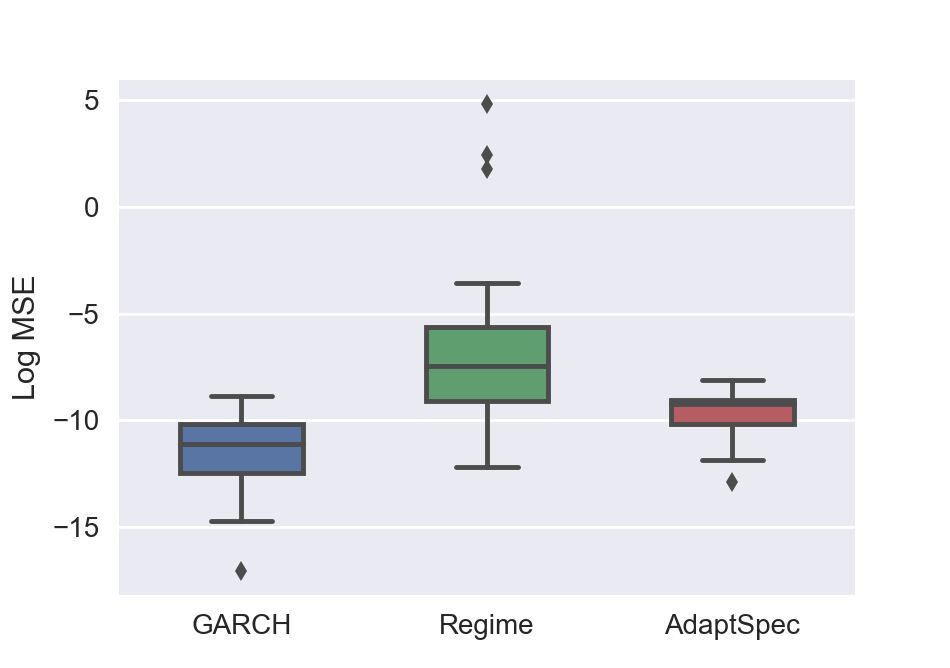}
\caption{GARCH Generated Data}
\label{fig:garch_mse}
\end{subfigure}
\begin{subfigure}{0.32\textwidth}
\includegraphics[width=\columnwidth]{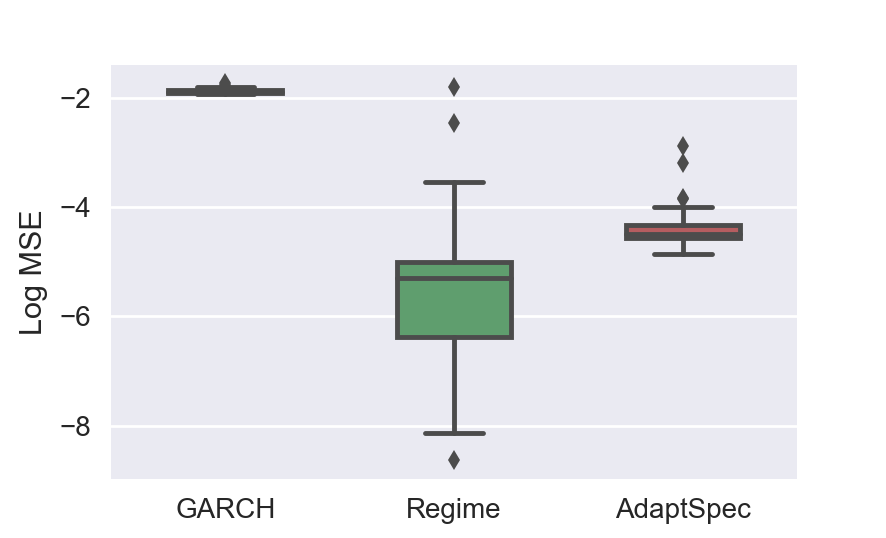}
\caption{Regime Generated Data}
\label{fig:regime_mse}
\end{subfigure}
\begin{subfigure}{0.32\textwidth}
\includegraphics[width=\columnwidth]{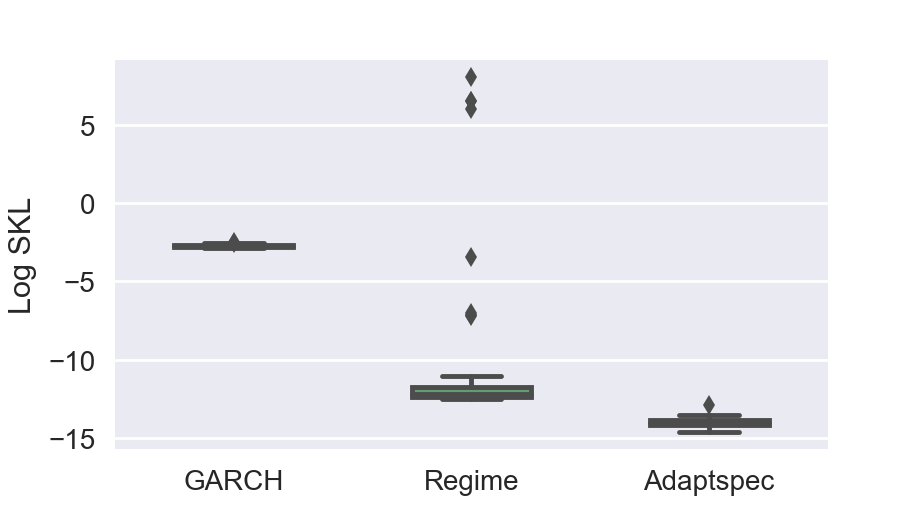}
\caption{AdaptSpec Generated Data}
\label{fig:adaptspec_mse}
\end{subfigure}
\caption{Boxplot of the $\log(MSE)$ for three estimators from 50 realisations generated from each respective process.}
\label{fig:simulation_box_plots}
\end{figure*}

Boxplots of the $\log(SKL)$ and $\log(MSE)$ for all three estimators and all three data generating models appear in Figure \ref{fig:simulation_box_plots}.  We chose to plot the log of these validation metrics, rather than the metrics itself, because the difference between the values of the $SKL$ and $MSE$ for three estimators is very large. 

As expected, when data are generated from a particular model, the estimates obtained from the method which assumes that particular model provide the best fit, (except in certain circumstance with the Regime model which will be discussed later). However, the plots also show that the estimates obtained from the AdaptSpec model when the data are generated from the GARCH or Regime models are always the next best.  For example \ref{fig:simulation_box_plots}, where the true model is a single GARCH model, which is the same as a Regime model where the number of regimes is equal to one, AdaptSpec outperforms the estimate obtained using the REGIME model. In other words, the improvement gained by using a flexible model, when flexibility is required, exceeds the loss of using a flexible model when flexibility is not required. 

\begin{figure*}[ht]
\centering
\begin{subfigure}{0.245\textwidth}
\includegraphics[width=\columnwidth]{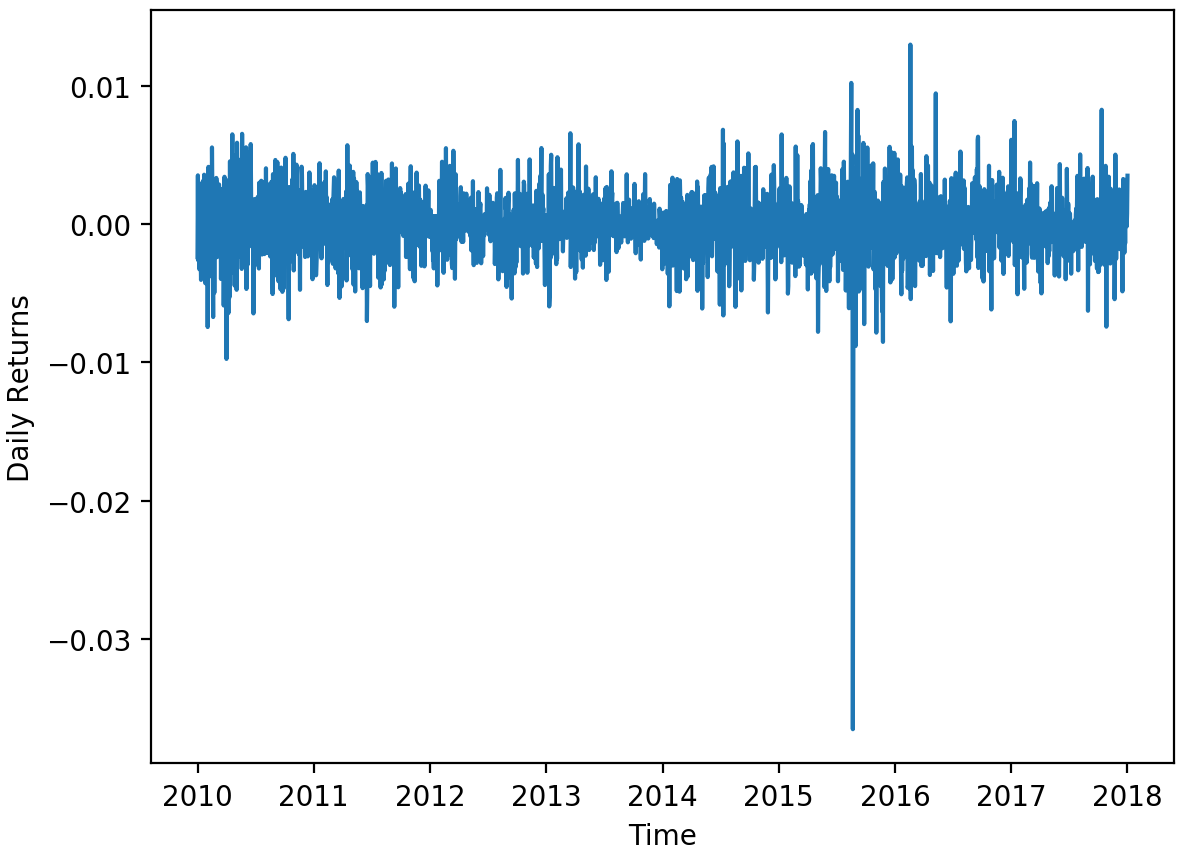}
\caption{GBP:USD Returns}
\label{fig:gbp_returns}
\end{subfigure}
\begin{subfigure}{0.245\textwidth}
\includegraphics[width=\columnwidth]{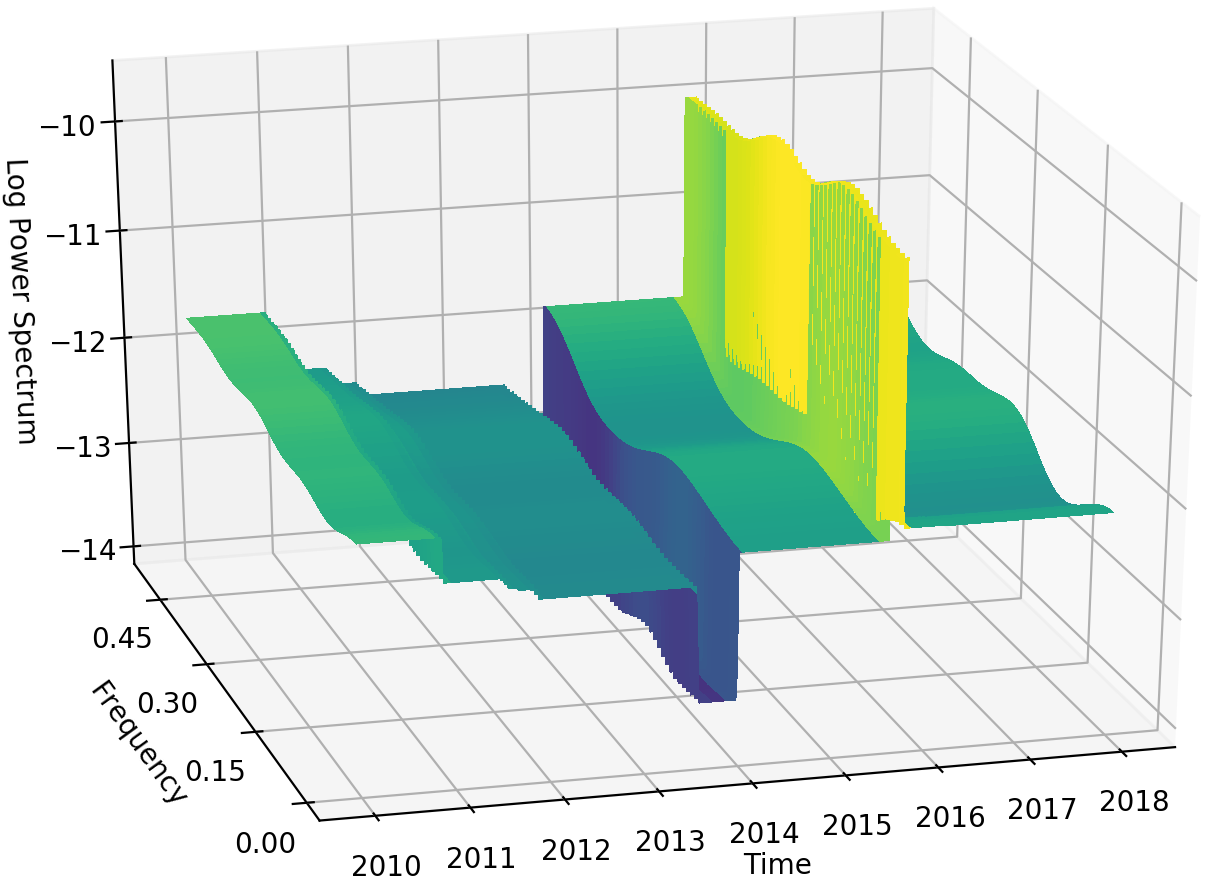}
\caption{GBP:USD Returns Log Spectrum}
\label{fig:gbp_squared_returns}
\end{subfigure}
\begin{subfigure}{0.245\textwidth}
\includegraphics[width=\columnwidth]{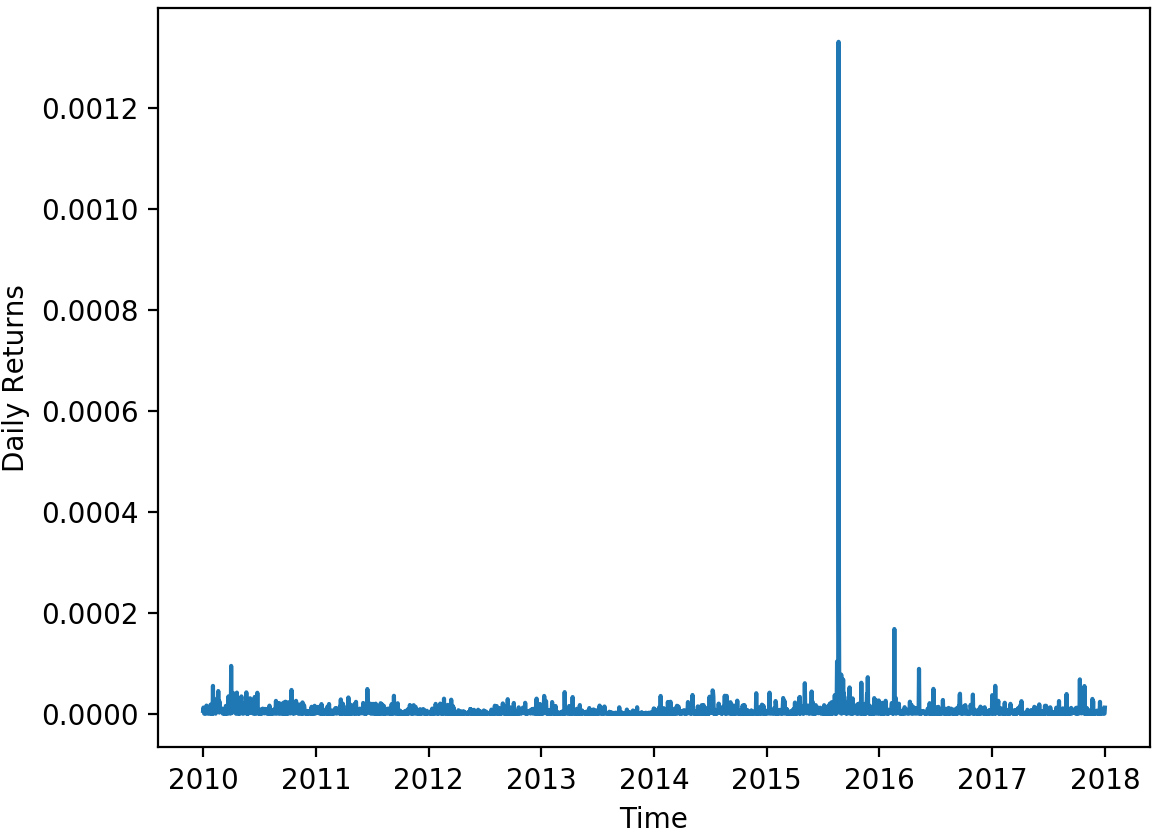}
\caption{GBP:USD Squared Returns}
\label{fig:gbp_spectrum}
\end{subfigure}
\begin{subfigure}{0.245\textwidth}
\includegraphics[width=\columnwidth]{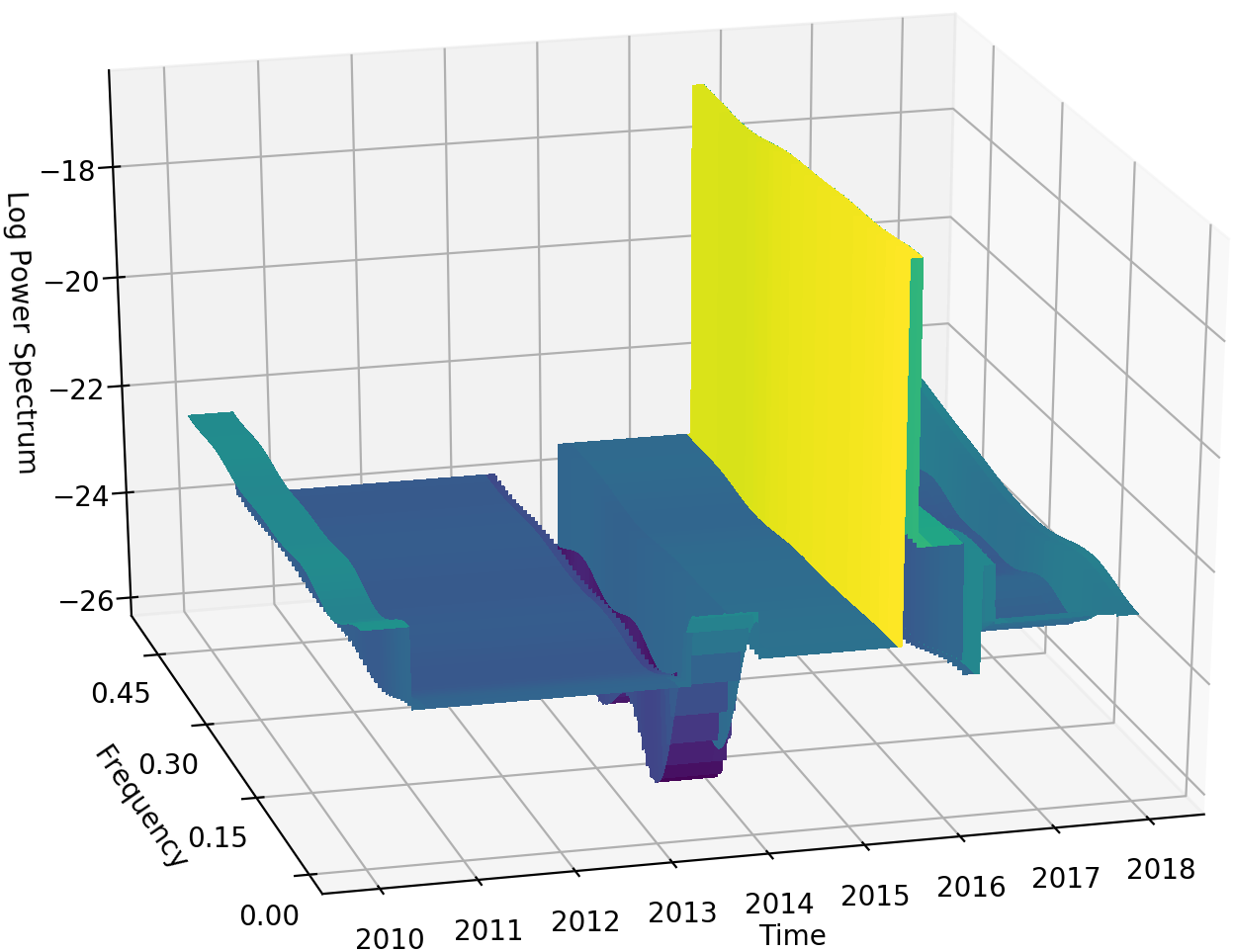}
\caption{GBP:USD Squared Returns Log Spectrum}
\label{fig:gbp_squared_spectrum}
\end{subfigure}
\caption{GBP:USD Returns and Log Spectra}
\label{fig:gbp}
\end{figure*}

\begin{figure*}[ht]
\centering
\begin{subfigure}{0.245\textwidth}
\centering
\includegraphics[width=\columnwidth]{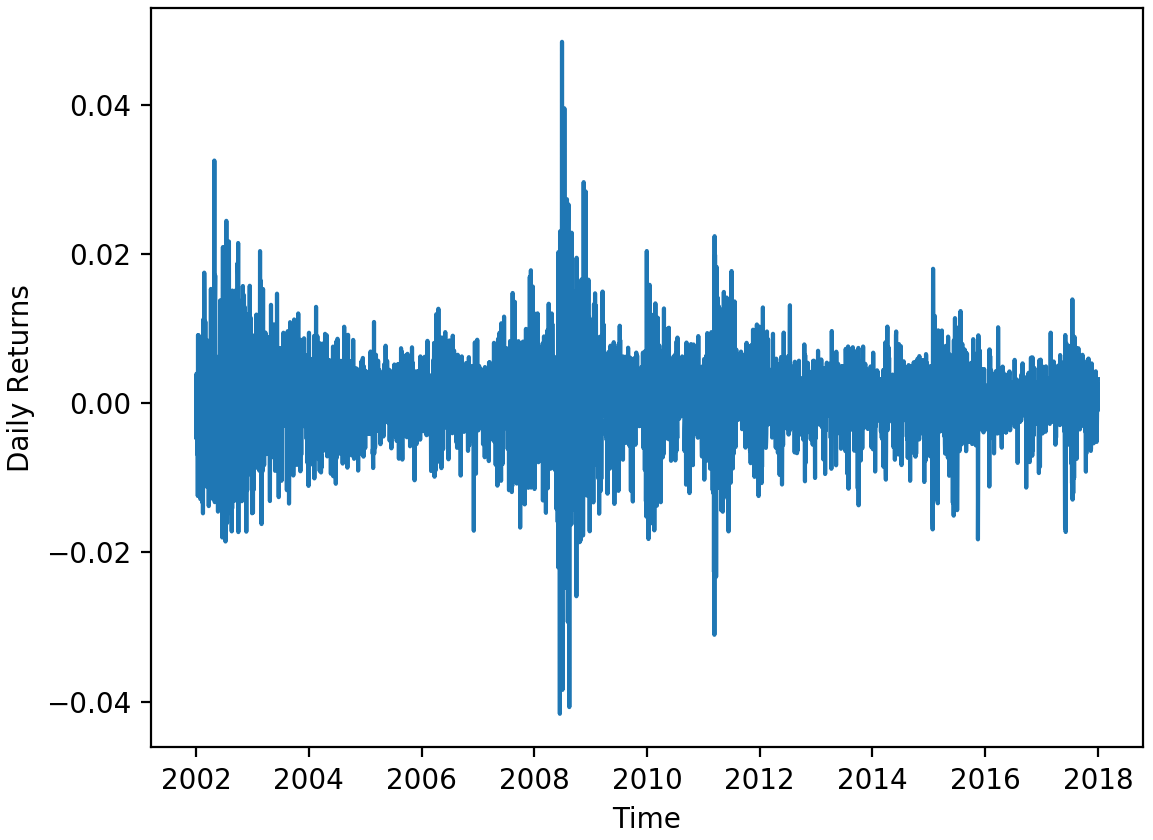}
\caption{NASDAQ Returns}
\label{fig:nasdaq_returns}
\end{subfigure}
\begin{subfigure}{0.245\textwidth}
\centering
\includegraphics[width=\columnwidth]{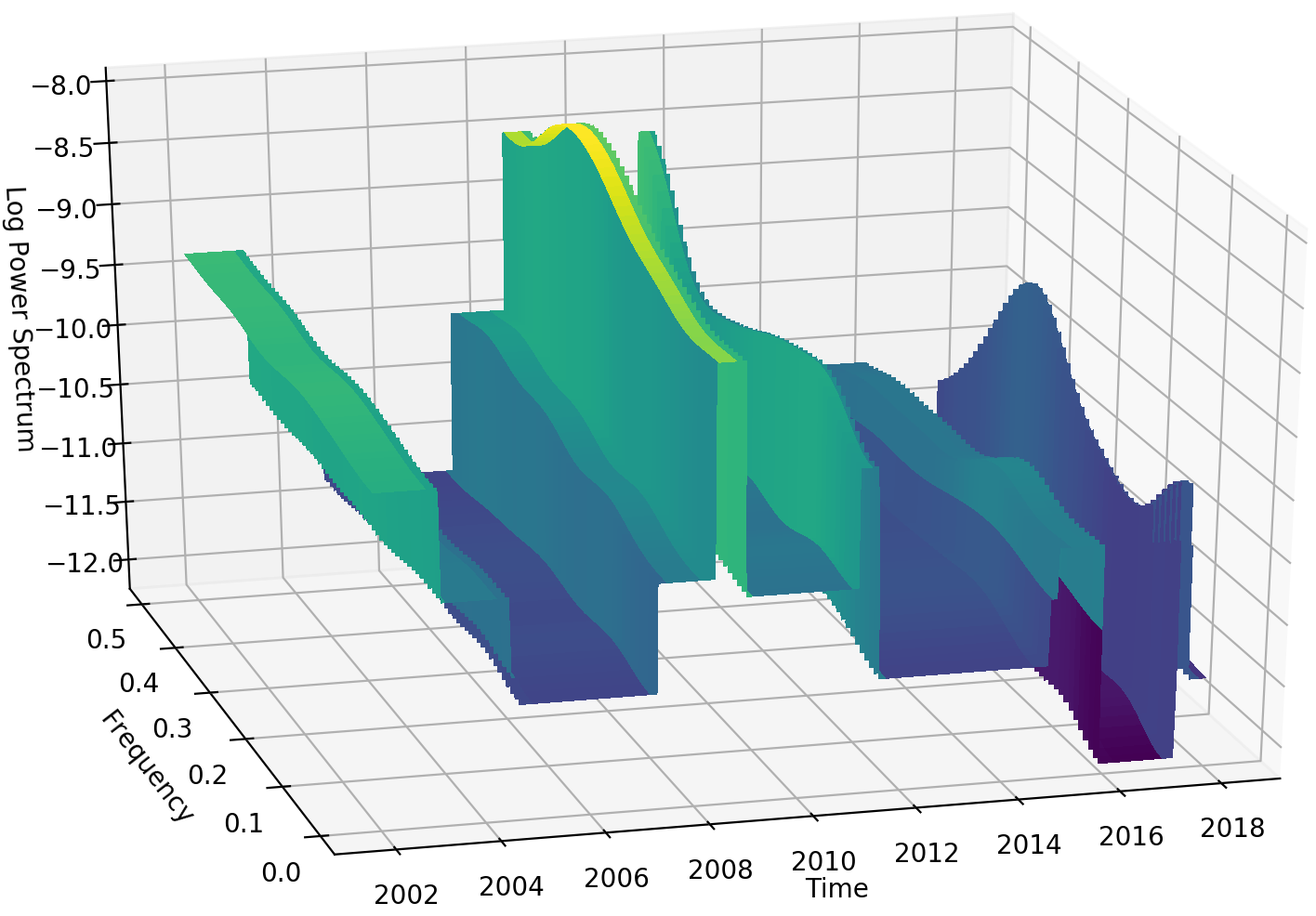}
\caption{NASDAQ Returns Log Spectrum}
\label{fig:nasdaq_spectrum}
\end{subfigure}
\begin{subfigure}{0.245\textwidth}
\centering
\includegraphics[width=\columnwidth]{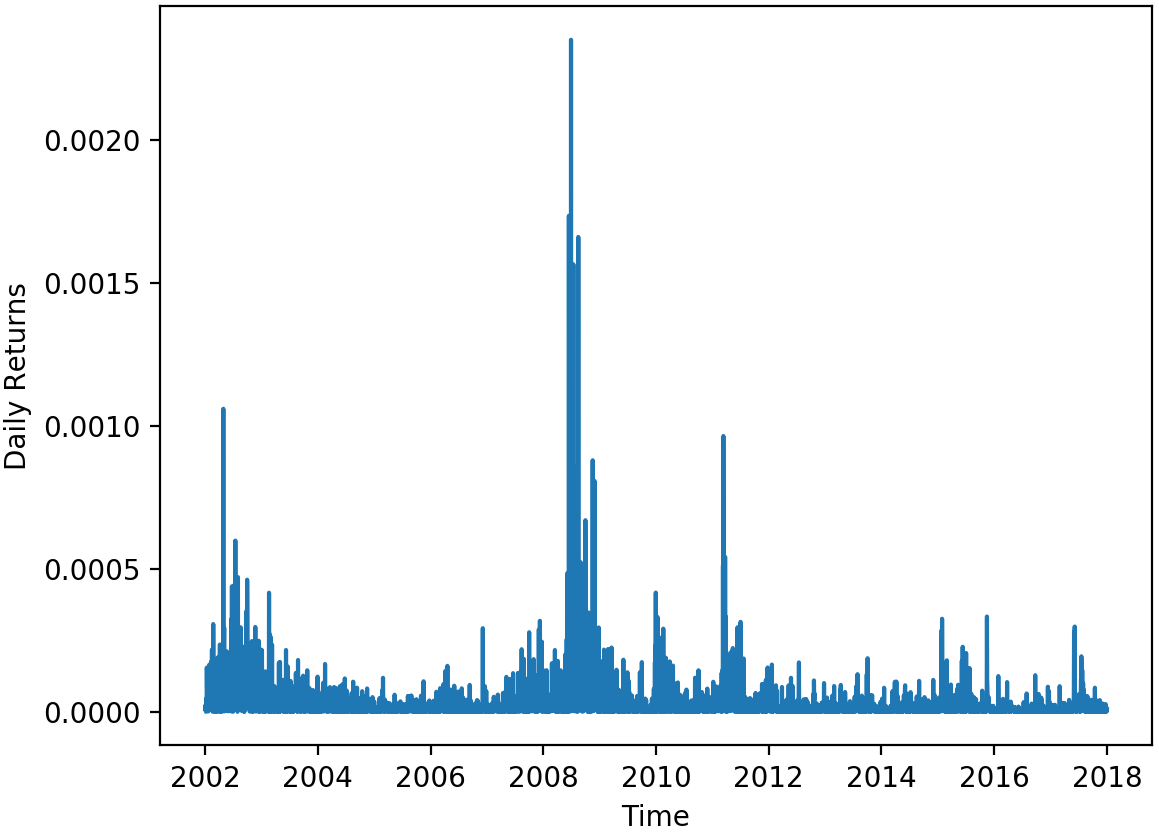}
\caption{NASDAQ Squared Returns}
\label{fig:nasdaq_squared_returns}
\end{subfigure}
\begin{subfigure}{0.245\textwidth}
\centering
\includegraphics[width=\columnwidth]{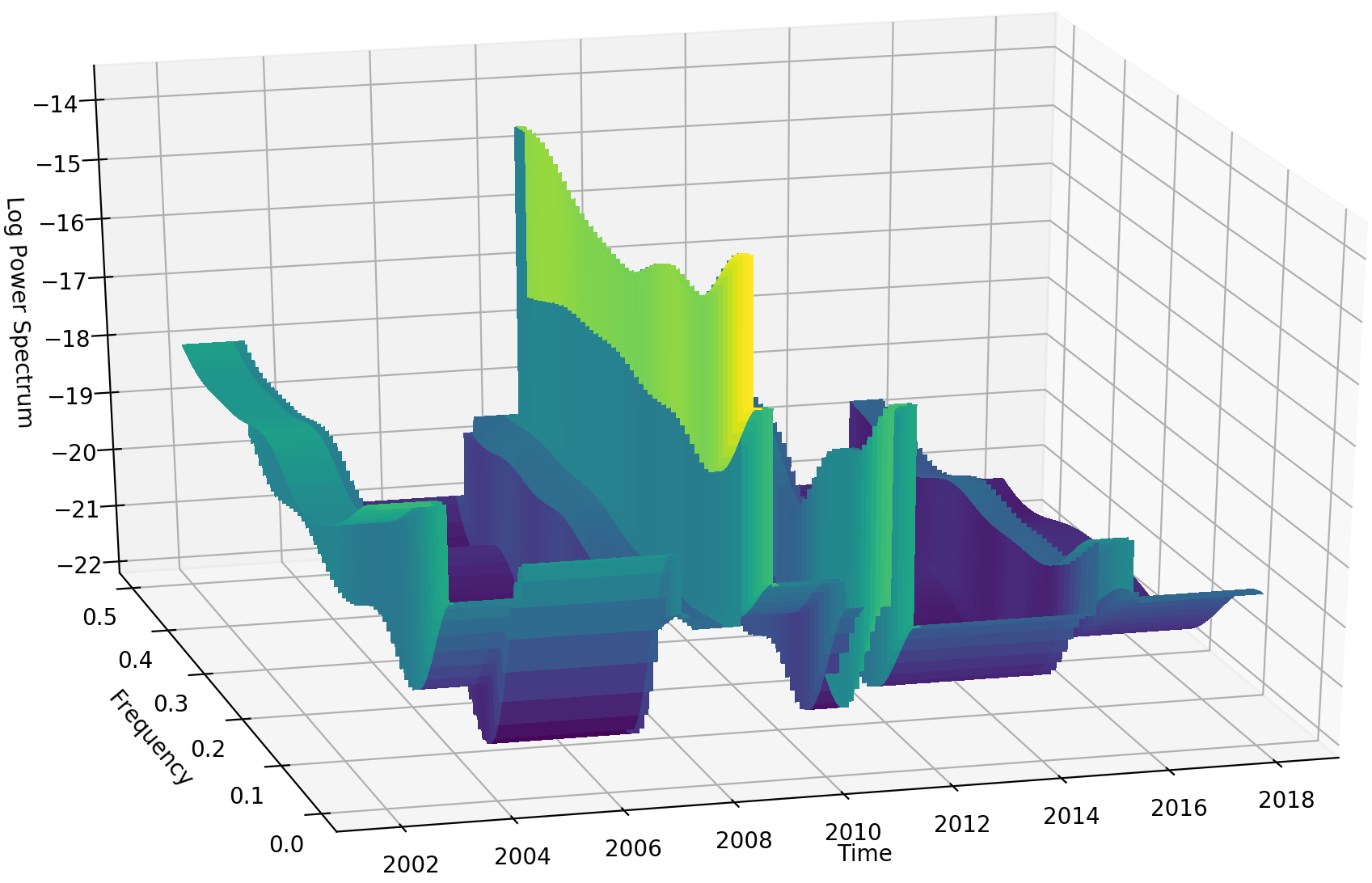}
\caption{NASDAQ Squared Returns Log Spectrum}
\label{fig:nasdaq_squared_spectrum}
\end{subfigure}
\caption{NASDAQ Returns and Log Spectra}
\label{fig:nasdaq}
\end{figure*}

The performance of the Regime model when data are generated from a GARCH model warrants further explanation. Our experience of using the model by \cite{Ardia2016}, shows that unless the true number of regimes is equal to the user-set number of regimes, results are highly variable. Part of the issue is an over-identification problem. If a single GARCH model is the truth but one estimates the spectrum using a regime switching model, where the number of regimes is greater one, then there are infinitely many different combinations which could recover the truth. While this should not necessarily present a problem with the estimated fit or prediction (as opposed to parameter inference), it does. This appears to be due to the fact that the probability of being in a particular regime can change abruptly on a  daily basis. These estimated probabilities in turn, are very sensitive to the specification of the particular type of GARCH model assumed to generate data in the different regimes. 

\begin{figure}[ht]
\vskip 0.2in
\begin{center}
\centerline{\includegraphics[width=\columnwidth]{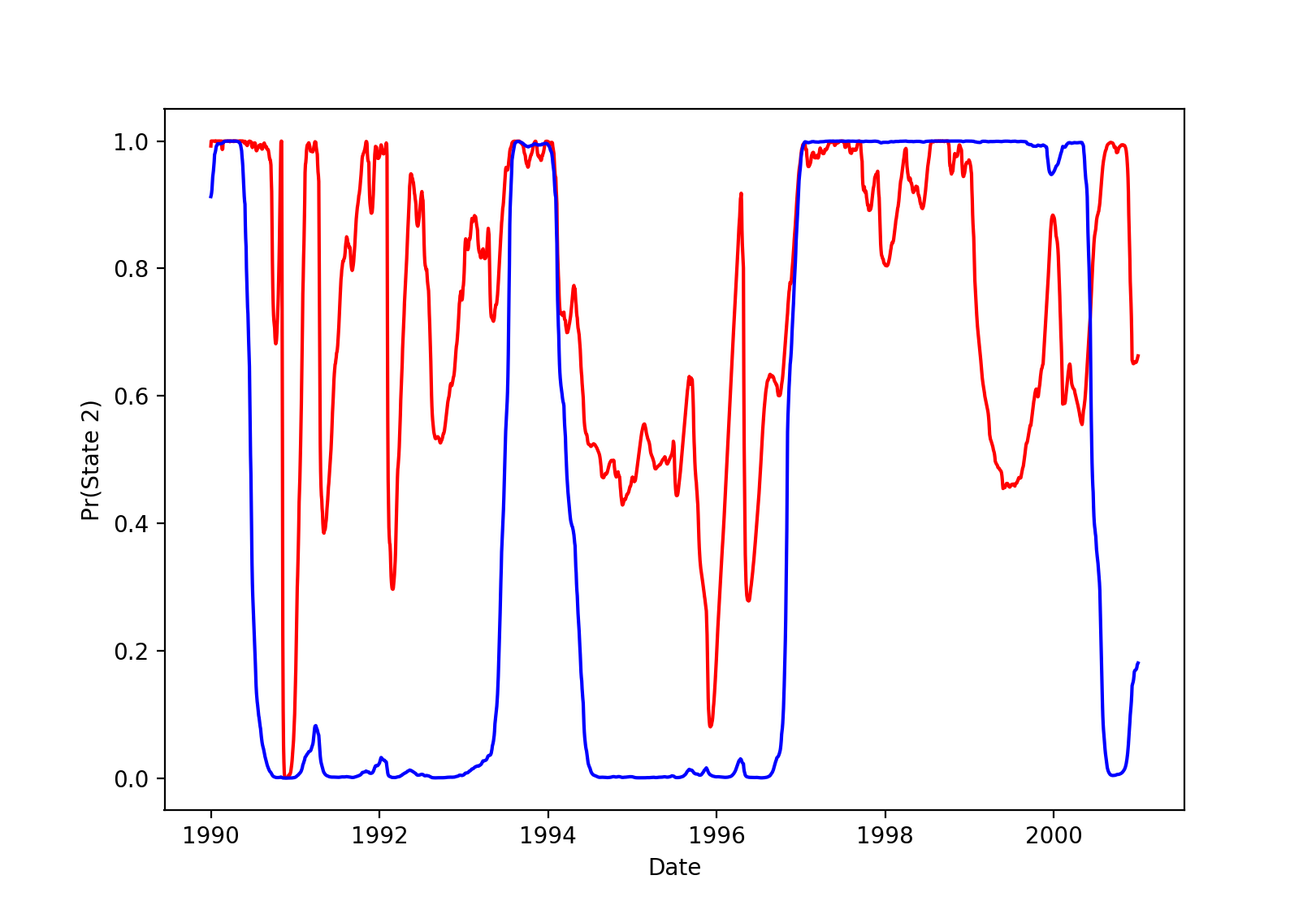}}
\caption{Estimated smoothed probabilities of the second regime}
\label{fig:probabilities_SMI}
\end{center}
\vskip -0.2in
\end{figure}

For example, we reproduced the "smoothed" probabilities obtained for the time series of the daily returns for the Swiss Market Index, which was analyzed by \citet{Ardia2016} These probabilities are the blue line in Figure \ref{fig:probabilities_SMI}, and are estimated using a Regime model assuming two GJR-GARCH processes. However, if we assume that the underlying data generating process for the regimes is a GARCH(1,1) rather than a GARCH(1,1) with a GJR variance specification \cite{Haas2004}, then we obtained the estimated smoothed probabilities given by the red line in Figure \ref{fig:probabilities_SMI}. The difference is striking.

These results also explain why AdaptSpec performs well across a range of data generating process; Adaptspec is a non-parametric model, so that by estimating the dependency in the frequency domain we avoid making any assumptions about the data generating process in the time domain.



\subsection{Real Examples: NASDAQ, GBP:USD}
\label{sec:experiment_real_data}
It is well known that the distribution of many financial assets are non-normally distributed, and exhibit volatility clustering. Whether this volatility clustering is evidence of long-range dependence in a stationary process, or attributable to non-stationarity is less clear.  In this section we attempt to answer this question by estimating the potentially time-varying spectrum of a financial time series' actual and squared returns. The time-varying spectrum of the actual return series is a non-parametric estimate of the evolution of the second moment of the return series' distribution, while the time-varying spectrum of the squared return series is a non-parametric estimate of the evolution of the fourth-moment. We choose the NASDAQ daily returns and the GBP:USD exchange rate daily returns from 2002-2018 and 2010-2018 respectively to demonstrate the technique.

Figure \ref{fig:nasdaq_returns} \ref{fig:nasdaq_spectrum} show the actual return series for the NASDAQ index and its estimated time-varying spectrum, while panels \ref{fig:nasdaq_squared_returns} \ref{fig:nasdaq_squared_spectrum} show the squared returns for the NASDAQ index and its corresponding estimated time-varying spectrum.  Figure \ref{fig:gbp} is an analogous plot for the GBP:USD exchange rate.

Figure \ref{fig:nasdaq} provides several insights into the stationarity and dependency of the NASDAQ returns. First, the series is definitely non-stationary.  The posterior mode of the number of locally stationary segments for the return series is 12. Second, it would appear that the market for the NASDAQ index is weak-form inefficient at several points in time. A weak-form efficient market is characterised by having zero autocorrelation in the first moment of the return distribution, and hence a flat spectrum. \ref{fig:nasdaq_spectrum} of Figure \ref{fig:nasdaq} shows several periods of time where the assumption of weak-form efficiency is violated, of particular note is the spectrum during the Global Financial Crisis (GFC) in 2008-2009, which shows a clear peak. \ref{fig:nasdaq_squared_spectrum} of Figure \ref{fig:nasdaq} shows that the volatility clustering is not removed even after accounting for non-stationarity. If non-stationarity accounted for volatility clustering then we would expect the locally stationary spectra of the squared return to be flat, however \ref{fig:nasdaq_squared_spectrum} shows that there is still strong positive correlation of the squared returns, as evidenced by the peak in power at low frequency for most of the time periods.

Figure \ref{fig:gbp} paints a similar picture for the GBP:USD exchange rate; the time series is clearly non-stationary, showing an overall increase in variability at the time of the Brexit vote with an accompanying dependency in the first moment of the series at that time, indicating violations of weak-form efficiency. However, the time varying spectral density of the GBP:USD squared returns as seen in \ref{fig:gbp_squared_spectrum} provides some interesting insights - distinguishing the behaviour of the GBP:USD's volatility with that of the NASDAQ Index. In particular, it indicates that non-stationarity drives the volatility clustering behaviour of the returns. This is clear because unlike the NASDAQ squared returns spectrum \ref{fig:nasdaq_squared_spectrum}, the GBP:USD squared returns spectrum \ref{fig:nasdaq_squared_spectrum} is predominantly flat within any candidate segment - suggesting that the larger non-stationary process is in fact piecewise stationary. 

\section{Conclusions}
\label{sec:conclusion}
Our experiments indicate that given a non-stationary data generating process, nonparametric models outperform parametric models, where the latter assumes a constant structure over time. Our simulations demonstrate that there is less estimation error in applying a flexible method such as AdaptSPEC to a parametric data generating process, than applying a parametric model to a non-stationary data generating process. For validation, we generate "ground truth" data in the spectral domain, and compare the resulting estimation from time domain models with spectral analysis techniques. The time series we generate after converting our ground truth spectrum into a time series strongly resembles many financial time series (such as the NASDAQ), and illustrates the need for flexible nonparametric models to capture the complex, non-stationary structure of the underlying time series.

\bibliography{example_paper}
\bibliographystyle{icml2019}

\end{document}